\documentclass[lettersize,journal]{IEEEtran}
\usepackage{amsmath,amsfonts}
\usepackage{algorithmic}
\usepackage{algorithm}
\usepackage{array}
\usepackage[caption=false,font=normalsize,labelfont=sf,textfont=sf]{subfig}
\usepackage{textcomp}
\usepackage{stfloats}
\usepackage{url}
\usepackage{verbatim}
\usepackage{graphicx}
\usepackage{cite}
\usepackage{changepage}
\usepackage{newtxtext,newtxmath}

\usepackage{multirow}

\usepackage{placeins}
\usepackage{booktabs, bm}

\usepackage{makecell}

\usepackage{listings}
\usepackage{rotating}

\begin{document}

\title{Gotta Go Fast: Measuring Input/Output Latencies of Virtual Reality 3D Engines for Cognitive Experiments}

\author{Taeho Kang, Christian Wallraven}

\markboth{Arxiv version, July~2023}%
{Shell \MakeLowercase{\textit{et al.}}: A Sample Article Using IEEEtran.cls for IEEE Journals}


\maketitle

\begin{abstract}
Virtual Reality (VR) is seeing increased adoption across many fields. The field of experimental cognitive science is also testing utilization of the technology combined with physiological measures such as electroencephalography (EEG) and eye tracking. Quantitative measures of human behavior and cognition process, however, are sensitive to minuscule time resolutions that are often overlooked in the scope of consumer-level VR hardware and software stacks. In this preliminary study, we implement VR testing environments in two prominent 3D Virtual Reality frameworks (Unity and Unreal Engine) to measure latency values for stimulus onset execution code to Head-Mount Display (HMD) pixel change, as well as the latency between human behavioral response input to its registration in the engine environment under a typical cognitive experiment hardware setup. We find that whereas the specifics of the latency may further be influenced by different hardware and software setups, the variations in consumer hardware is apparent regardless and report detailed statistics on these latencies. Such consideration should be taken into account when designing VR-based cognitive experiments that measure human behavior.
\end{abstract}


\section{Introduction}
\IEEEPARstart{T}{he}
idea of utilizing naturalistic stimuli in cognitive experiments is increasingly gaining traction, and its importance has been recognized in an increasing number of studies~\cite{hasson2004intersubject, degelder2018virtual, bacha2018drama, jaaskelainen2021movies, finn2020idiosynchrony, salmi2014posterior, schmuckler2001ecological}.
3D environments such as Virtual and Mixed Reality (VR/MR) can provide an excellent platform for implementing experimental paradigms where immersive, interactive and naturalistic stimuli presentation is desired~\cite{tarr2002virtual, pan2018and, de2018virtual}.	
Especially for VR, behavioral and cognitive investigative experiments performed in virtual reality have the advantage of being able to control and manipulate environmental variables that in real-life settings would be nearly impossible to control
~\cite{stephens2009situational, pan2018and}.
Possibly due to this, virtual reality has seen increased utilization in the area of behavioral and cognitive investigations, from simple behavioral experiments~\cite{rosenberg2013virtual}, to neuroimaging studies~\cite{mueller2012building,}, to even timing sensitive studies involving physiological signal measurements such as EEG~\cite{brouwer2011eeg, i2012using, faller2017feasibility, kuziek2020real, tarng2019towards, jang2022decoding, tarrant2018virtual, bayliss2000virtual}.	

Due to the nature of cognitive processes of interest, behavioral and cognitive studies investigating timing-critical brain processes have been historically sensitive to latency in experimental hardware and software~\cite{plant2003choice, garaizar2014measuring}.
It has been suggested, however, that specialized behavioral input devices may not be as crucial for even time-critical experiments, as the variability from human behavior itself is generally larger in scale than the input lag occurring from individual hardware devices~\cite{damian2010does}. For ease of experimental equipment acquisition that may in turn be relevant to the easy replication of studies, usage of adequately performant consumer hardware may be preferable to limited costly specialized equipment for behavioral input. 

Nonetheless, especially in studies measuring time-sensitive behaviors, there is importance in measuring expected latency in hardware and software setups used for experimental paradigms. Wimmer et al.~\cite{wimmer2019latency} used opto-couplers to measure latency of 36 different serial input devices connected to a Raspberry Pi device and formed probability distribution models for each of the devices, and reported different input latency distributions per device; suggesting the need for measuring input latency levels in interactive experimental setups that make use of serial device based user input device.
Furthermore, due to higher graphical computation requirements than conventional displays arising from not only generally higher refresh rates but also other factors such as needing to render twice for stereo vision, the current state of VR hardware suffers from latency greater than of conventional user interface devices~\cite{elbamby2018toward}.

A final point of consideration in this context concerns the use of higher-level APIs for generating three-dimensional, interactive environments that afford realistic levels of sensory realism and interactivity. While it is possible to create well-controlled low-level stimuli with relative ease in computer graphics languages, the amount of work necessary to create environments which, for example, contain objects that interact with each other in a physically-realistic fashion from scratch is beyond the capabilities of standard cognitive and behavioral research labs. For this reason, many researchers have increasingly turned to 3D game engines for creating such environments. One drawback of this development is that these engines offer only a reduced degree of control over their timing internals given that much of the behind-the-scenes calculation remains hidden from the API user (examples include the calculation of graphic primitives, the determination of collisions in physics-aware simulations, etc.). This raises the question of how much timing accuracy and precision is possible in game engine simulation programming environments. In this context, Wiesing et al.~\cite{wiesing2020accuracy} have measured stimuli duration and onset measurements in Unreal Engine with a dedicated response pad and reported increased average latency, compared to dedicated cognitive experiment software such as Psychopy and Psychtoolbox. While Unreal Engine as a serious 3D engine has been used for VR based behavioral experiments~\cite{shapcott2022domevr}, possibly due to the relative ease of implementation in comparison to the former, Unity Engine has been seeing increased applications in cognitive experiments~\cite{ju2019brake, nezami2021westdrive, ju2020acoustic, brookes2020studying, vortmann2019eeg, quintero2022excite, kuziek2020real}. While they are suited for similar purposes, due to differences in implementation detail Unity and Unreal Engine often exhibit different behaviors even when the same effect is intended, especially in frame and I/O related latency performances~\cite{vsmid2017comparison}.

In light of these considerations, to ultimately implement and execute experiments investigating brain processes in a naturalistic VR environment, we deem it worth investigating the expected latency values for hardware and software setups that would (commonly) be used in VR-based behavioral experiments. In this study, we aim to achieve this by utilizing a measuring apparatus with oscilloscopes, as well as a bare-bone experimental paradigm implemented in two widely-used VR capable 3D engines, Unity Engine and Unreal Engine. In the bare bone paradigm, we create stimulus onsets that send trigger codes to the measuring apparatus before the actual displaying of stimuli is performed in the Head Mounted Device (HMD), and measure latency between the the onset code and the actual pixel change in the HMD. Furthermore, we measure the latency between a physical input action on consumer-level user interface hardware (keyboard) that can be used for behavioral response, and the registration of the input in the 3D engines. Lastly, we measure the latency between the physical input action and the pixel change from the resultant feedback code execution.

	\section{Materials and Methods}
	\subsection{Experiment Design}
	
	\begin{figure}[htbp]
		\centering
		\captionsetup{justification=centering}
			\includegraphics[width=0.45\textwidth]{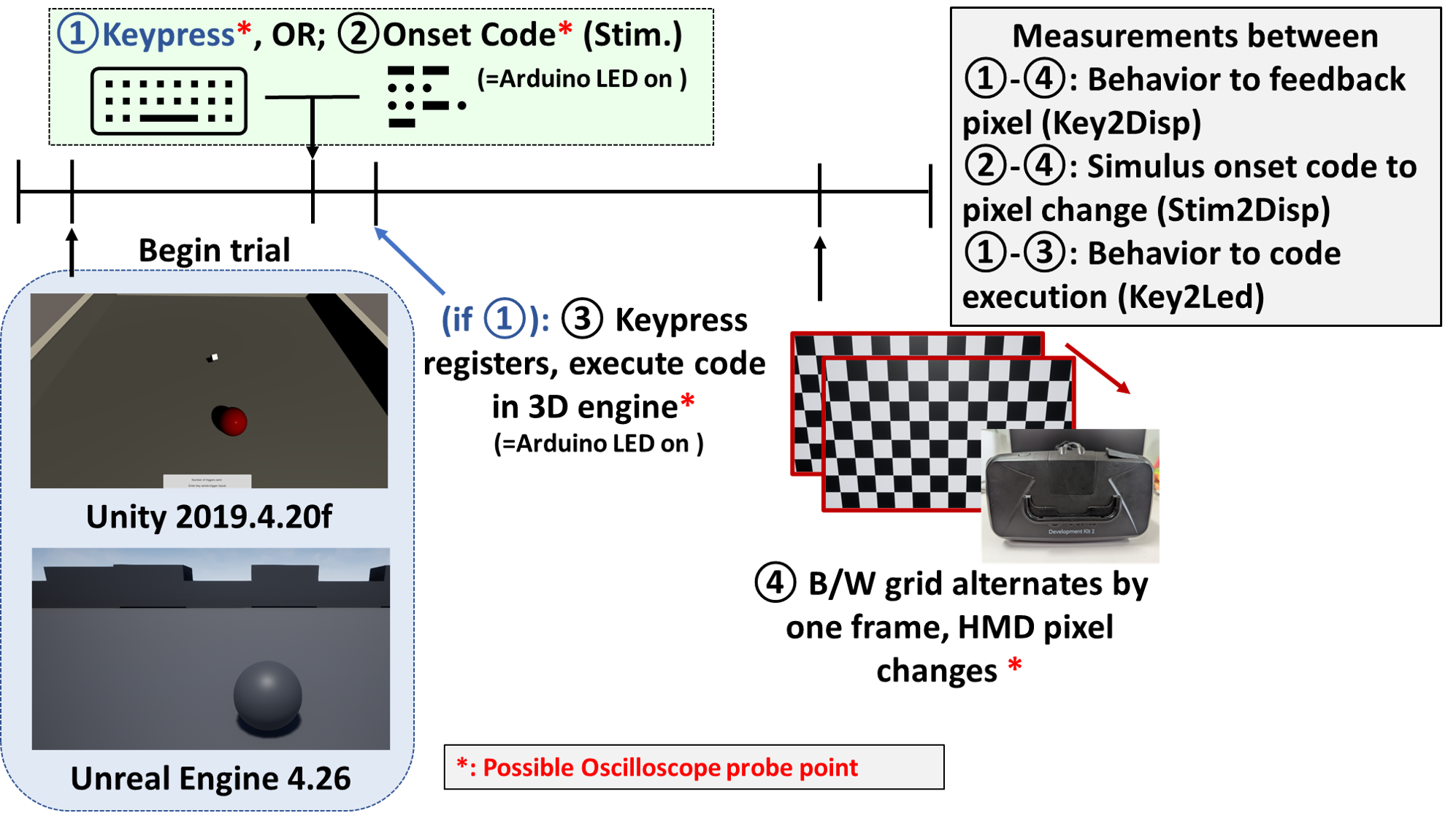}	
		\caption{Diagram of the experimental data collection paradigm trial design.}
		\label{fig:c4_diagram_trial}
	\end{figure}
We were interested in the measurement of the following events: 1) latency between stimulus onset code execution in the 3D engine and the actual HMD pixel changes as the stimulus was presented (Stim2Disp), 2) latency between participant behavioral response by a keypress and the code execution performed immediately upon the register of the key event in the 3D engine (Key2Led), and 3) latency between participant behavioral response by keypress and the pixel change in HMD caused by the 3D engine code that presents a visual feedback stimulus upon the key event register (Key2Disp). To measure these events, we implemented experimental paradigms capable of measuring these events in both Unity Engine and Unreal Engine, as can be seen in Figure~\ref{fig:c4_diagram_trial}. 

We decided to measure both 3D Engines as their implementations differ, as well as the scripting implementation language for user code: Unity utilizes C\# for this, whereas Unreal uses C++. Both engines have seen usage in cognitive experimental designs in 3D environments (see Introduction).

The 3D VR experiment environment consists of a basic 3D spherical object that can move around based on the user's input. Upon a stimulus onset code or recognition of a participant's behavioral response by a keyboard, a chess-board grid of black and white covers the entire screen for one frame, then with the colors inverted for another frame. The experiment code embodies a bare-bone basic form of cognitive experimental paradigm using 3D Engines, and we send programmatic triggers upon stimulus onset and behavioral input registration, as one would for experiments requiring high temporal resolution like in EEG or other cognitive experiments that involve some form of time series physiological measurement. We measure the latency of the above 3 scenarios (Stim2Disp, Key2Disp, Key2Led) by sending programmatic markers to Arduino upon stimulus onset code execution, and behavioral response registration in the 3D engine, which triggers an LED. Furthermore, we use pressure sensors and photodiodes connected to the Arduino board to have quantifiable measures of when participant behavior and stimuli display happen in the real world.
	
\subsubsection{Unity-specific setup}
For Unity, we implemented the paradigm in Unity Engine version 2019.4.20f.
Following Unity Engine's manual on order of execution for event functions (see\url{https://docs.unity3d.com/Manual/ExecutionOrder.html}), the FixedUpdate() function handles ticks on computations focused on the physics engine calculations, and may render more than once per rendered frame depending on the computation load and settings, while the Update() function ticks per every frame that is rendered (after FixedUpdate() calls are complete). The events are called serially, and between the Update() call and the actual rendering calls of the scene on display there are several other calls; as such, in the interest of sending the trigger for stimulus onset in experimental behavior measurement database as close to the onset of the actual stimulus in the display, it is preferable to send the marker code sometime after the Update() call but before the actual display rendering process. We achieve this by calling a coroutine that waits execution of sending stimulus onset triggers until the rendering computation is complete, but before the displaying is performed:


 \begin{figure}
    \begin{lstlisting}[frame=single]
    private IEnumerator HandleTriggers(GameObject grid1, GameObject grid2)
    {	
	    // because the first EOF is still on the same frame, wait
	    yield return new WaitForEndOfFrame();   
	    // send LED signal RIGHT BEFORE FRAME RENDERING
	    duino_port.Write(sevent_keypress, 0, 1); 
	    yield return new WaitForEndOfFrame(); 
	    // the stimuli is two frames of visual stimuli
	    // as such, disable the first frame and enable the other
	    grid1.SetActive(false); 
	    grid2.SetActive(true); 
	    grid2.SetActive(false);
	    yield return null;

    }	
	\end{lstlisting}
 \caption{Code for sending trigger to Arduino in Unity}
 \end{figure}

Furthermore, as the FixedUpdate() call executes at the beginning of the game tick and executes at a higher rate, it is preferable to process keyboard input events (i.e. behavioral response) in there. For Key2Led and Key2Disp events, the function handling keyboard input events can also call for the co-routines to send markers subsequently:
	
 \begin{figure}
 \begin{lstlisting}[frame=single]
	private void FixedUpdate()
	{	
		// updating the speed for the cueball
		mvVec3.Set(mvVec.x, 0.0f, mvVec.y);
		// actual position update and physicis 
		// are handled internally by unity
		rb.AddForce(mvVec3 * mvspeed);	
		// check update for keypresses in fixedupdate 
		// for fastest response possible
		if (keypressed)
		{		
			mrklsl_keypress.Write("Keypress");
			// handle stimulus triggering
			HandleTriggers(gridquad_1, gridquad_2);		
		}
		// reset keypress flag
		keypressed = false; 
	}
	\end{lstlisting}
    \caption{Unity code for sending trigger on behavioral response}
\end{figure}
\subsubsection{Unreal-specific setup}
For the Unreal Engine, we implemented the paradigm in version 4.26. 
Unreal Engine logic ticks are separated in tick groups (PrePhysics, DuringPhysics, PostPhysics, and PostUpdateWork) that are serially run as per the documentation (\url{https://docs.unrealengine.com/5.1/en-US/actor-ticking-in-unreal-engine/}). Processing of user input is handled in the PrePhysics segment of the ticking, and as such binding the input of specific keys to a method that sends trigger is sufficient. By calling stimulus onset triggers in a code that is executed on the PostPhysics or PostUpdateWork segment, we can also set it to be as close to the timing of the actual display as possible. To ensure tweak Unreal Engine for optimal performance, several project settings were changed in addition: First, in the Rendering->VR settings, Instanced Stereo was enabled while mobile HDR was disabled, as per recommendations by ~\cite{wiesing2020accuracy}. Second, the following console variables were were changed: R.GTSyncType to 1, R.Vsync to 0, and R.OneFrameThreadLag to 0. R.GTSyncType determines which thread in game processes sync to: 0 if they sync with the rendering thread, if 1 they sync to the RHI(render hardware interface=d3dx or opengl) thread. As per the Unreal documentation syncing to the RHI thread helps with input latency, so we set it as 1. VSync renders the frames at the pace at which the display device is capable of, but it often leads to more dropped frames when enabled than otherwise~\cite{raaen2015measuring}. When OneFrameThreadLag is enabled, the graphics drivers keep the game thread from processing further than one frame worth of computations than what is currently being displayed. We deemed this undesirable as our purpose was to minimize lags stemming from computations not being ahead enough, along with minimization of the input latency.
	
 \begin{figure}
 \begin{lstlisting}[frame=single]
	void Aball3d_426Ball::TriggerStim() 
	{	
		FTimerHandle dispTrigger;
		// send trigger to arudino
		if (!WriteFile(hSerial, &ledcode, 1, &bytesw, 0))
		{
			//In case it don't work get comm error and return false
			std::cout << "Writefile failed" << std::endl;
		} //!Stim1TimerFlag
		
		GridPlane1->SetVisibility(true);
		Stim1TimerFlag = true;	
		lastRender = FDateTime::Now();
		lastRender64 = lastRender.ToUnixTimestamp();
		lastRender32 = lastRender.GetMillisecond();
	}
	\end{lstlisting}
 \caption{Code for sending Arduino trigger on Unreal}
\end{figure}	

\subsection{Measuring apparatus}
\begin{figure}[htbp]
		\centering
		\captionsetup{justification=centering}
			\subfloat[Circuit diagram]{
				\includegraphics[width=.45\textwidth]{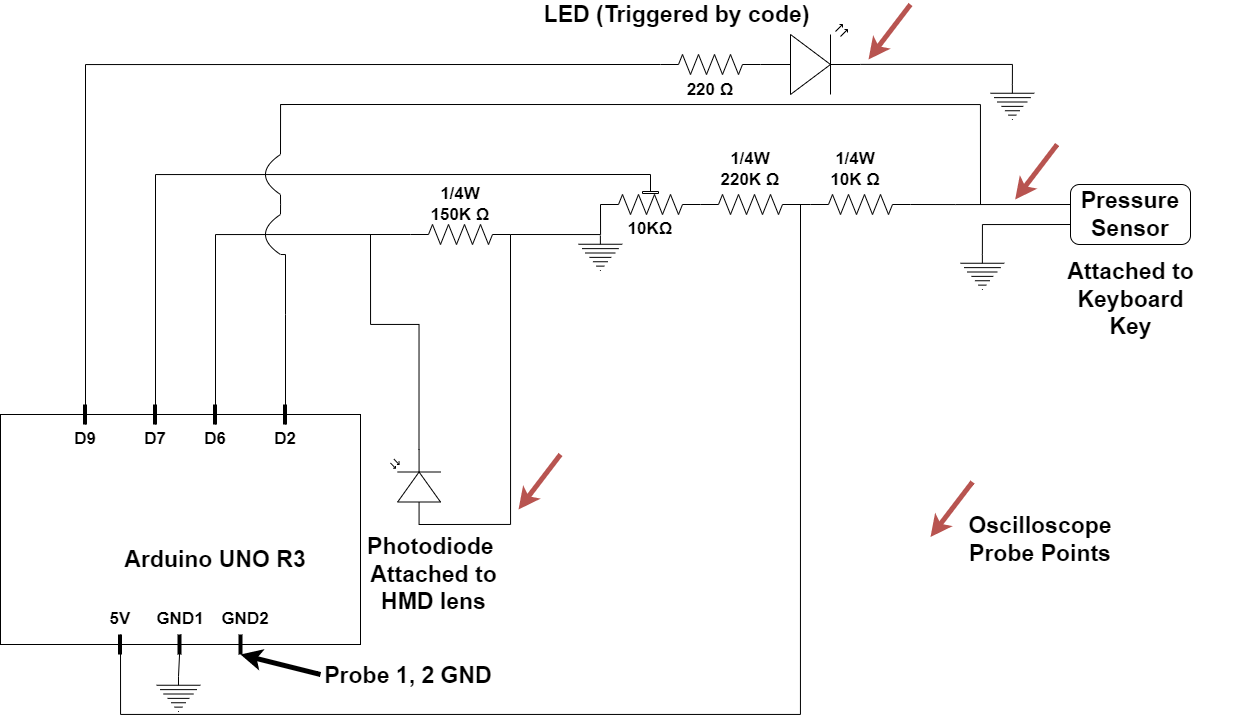}
				\label{subf-c4-diagram}
			}
			
			\subfloat[Setup]{
				\includegraphics[width=.45\textwidth]{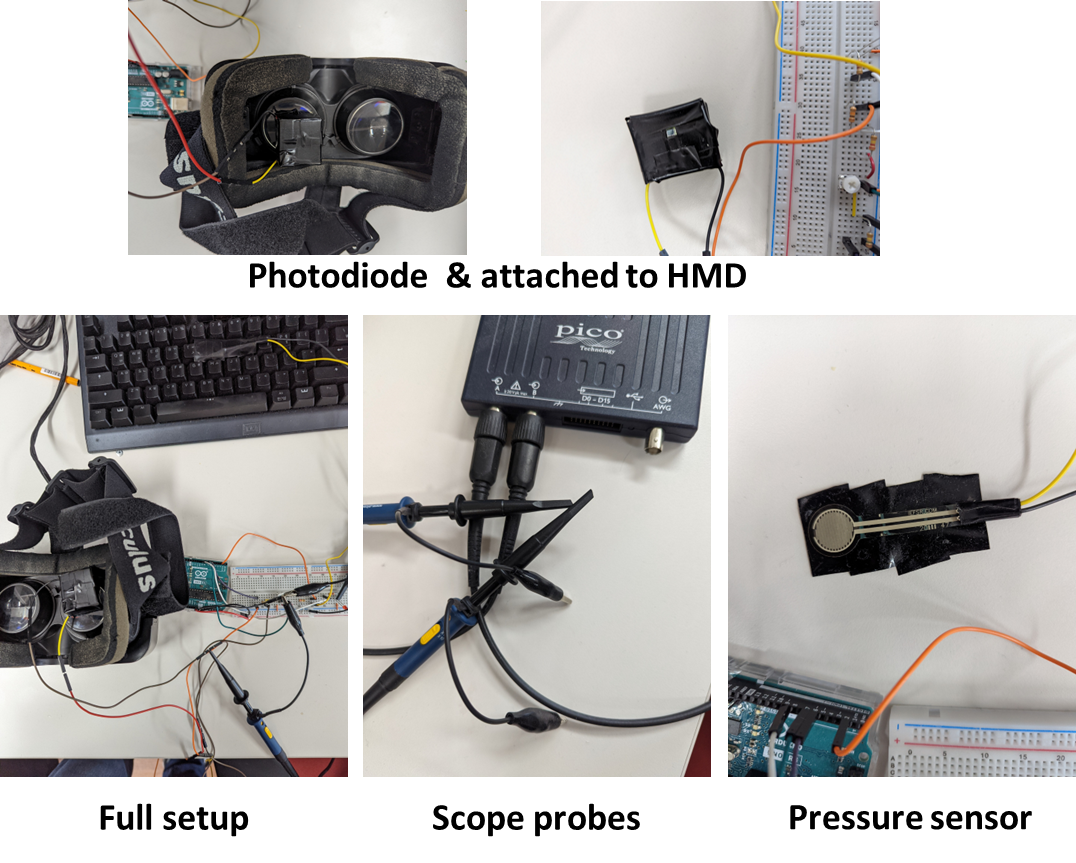}
				\label{subf-c4-setup}
			}
			
		\caption{Circuit diagram of the latency measuring apparatus, setup pictures}
		\label{fig:c4_diagram_setup}
	\end{figure}

To measure timings of 1) behavioral response onset, 2) stimulus onset code execution, 3) feedback code execution in response to behavior, and 4) pixel changes on the HMD as precisely as possible, a circuit apparatus using that can bee seen in Figure~\ref{fig:c4_diagram_setup} was implemented. 
The inspiration for the circuit board was based on a schematic from class material in Aachen university's system design course~\cite{borchers2022designing}. Specific components of the circuitry included an Arduino Uno Rev.3, a BPW-34  photosensitive diode developed by Vishay Semiconductors, a pressure sensor FSR402 developed by Interlink Electronics. For registering the behavior response, a Wooting One keyboard developed by Wooting was used. For running the 3D Engine based experimental paradigms, a Windows-10 based computer running on AMD's 5900x CPU with Nvidia RTX 3090Ti was used.

As the actual display of the VR environment, an Oculus DK2 headset from Meta Inc. was used, in which the photodiode was attached to next to the display.
An USB-based oscilloscope developed by Pico Technology (Picoscope series 2205A) with two probes was used for measuring the changes in voltage. The oscilloscope's sampling frequency was set too 240kHz. 
The ground clamps of both probes were connected to the ground pin cable of the Arduino board. As the Oculus DK2's refresh rate is at 75Hz, we band-pass filtered the probe connected to the photodiode to [60 80]Hz. Sample collection length per trial was set to 200ms, with 20ms pre-trigger and 180ms post-trigger. For the Stim2Disp measurements, the first probe was clamped to the LED diode that would be toggled on and off by the stimulus control code in the 3D engine, while the second probe was clamped to the cable connected to the HMD-attached photodiode. The scope data collection trigger was set to a rising threshold of 1.5V for Unity and 115mV for Unreal with a hysteresis of 5.87\% on the first probe. All scope measure trigger thresholds were set manually set after trial and error for catching the events of interest, and the difference in threshold was made due to the probe attenuator settings changed to different levels between the two sets of measurements, the difference in thresholds, however, did not interfere with the trigger adequately capturing the point where the event was occurring. This was verified after measurements by visually inspecting the probe waveforms for peaks from LED and keypress actions. In Key2Disp and Key2LED measurements, the first probe was clamped to the cable attached to the pressure sensor attached to the keyboard. Here again, due to selecting different probe attenuator settings, the probe trigger threshold was set to 450mV rise with 2.44\% hysteresis for Unreal, and a 4.7V rise for unity. The second probe was connected to the LED in Key2LED measurements, and to the photodiode in Key2Disp measurements. The Arduino would be connected to the experiment PC via USB, through which LED trigger communications would be sent from the 3D engines via serial communication.
For each latency event of interest, we made at least 300 repetitions of the measurement in order to collect sufficient sample size.
	
\subsection{Data processing}
Data preprocessing and analysis were performed with Matlab 2021b by Mathworks Inc.. As the scope sampling rate was rather high considering our time epochs of interest, data was first downsampled to 20kHz. As the data epochs were temporally zero-centered to the triggering event of the first probe, the timing of the events of interest on the second probe (photodiode voltage change, LED power on) had to be found by peak detection, as the onset of events of interest would result in significant changes in the probe voltage. In photodiode measurements this meant voltage troughs that were far greater than the baseline pixels (as the black and white grids would trigger a greater change in the luminosity of the display, leading to greater voltage changes). For all measurements on the second probe, as we were looking for the latency for the onset of the event of interest, finding the timing of only the first significant peak detected was necessary. Finding the position of the peaks was performed with the findpeak() function provided in the Signal Processing Toolbox of Matlab. The resultant peaks were plotted and manually inspected for enough number of trials (>100) in each condition to ensure the function was performing as desired. Once the timing of the events of interest were found, we calculated latency for the three events of interest.

	
\section{Results}
\begin{figure}[htbp]
			\subfloat[Stimulus onset - HMD pixel change(Stim2Disp), Unity 2019.4.20f]{
				\includegraphics[width=.45\textwidth]{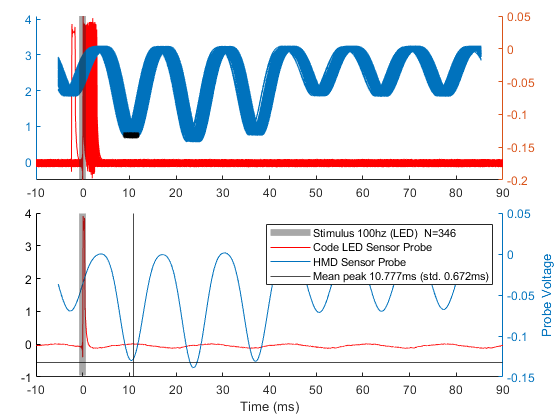}
				\label{subf-c4-stim2disp100-unity}
			}
			
			\subfloat[Stimulus onset - HMD pixel change(Stim2Disp), Unreal Engine 4.26]{
				\includegraphics[width=.45\textwidth]{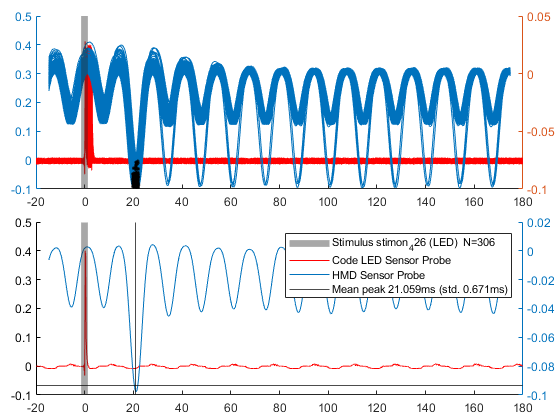}
				\label{subf-c4-stim2disp-unreal}
			}		
		\caption{Latency measured between stimulus onset marker execution code and actual stimuli onset pixel change in HMD (Stim2Disp), for both Unity and Unreal Engine. Higher plot in each set of plots shows all measurements from the two probes superimposed, along with the delayed response peaks determined by Matlab's findpeak() function. Lower plot shows the average waveforms, as well as the average of the automatically detected first onset of peaks in the second probe.}
		\label{fig:c4_stim2disp_result}
	\end{figure}

\begin{figure}[htbp]
		\centering
		\captionsetup{justification=centering}
			\subfloat[Keypress Behavior - HMD pixel change(Key2Disp), Unity 2019.4.20f]{
				\includegraphics[width=.45\textwidth]{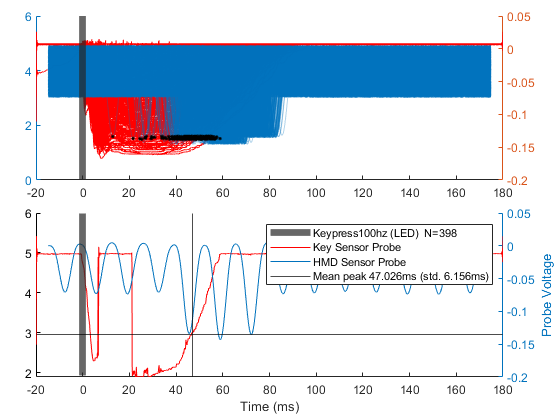}
				\label{subf-c4-key2disp100-unity}
			}
			
			\subfloat[Keypress Behavior - HMD pixel change(Key2Disp), Unreal Engine 4.26]{
				\includegraphics[width=.45\textwidth]{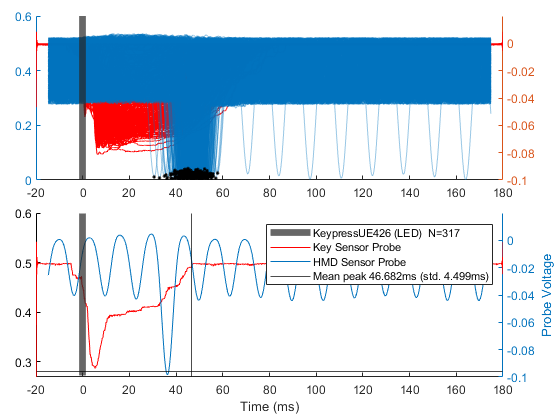}
				\label{subf-c4-key2disp-unreal}
			}		
		\caption{Latency measured between keyboard press behavior and feedback stimuli onset on HMD pixel in 3D engine (Key2Disp), for both Unity and Unreal Engine. Higher plot in each set of plots shows all measurements from the two probes superimposed, along with the delayed response peaks determined by Matlab's findpeak() function. Lower plot shows the average waveforms, as well as the average of the automatically detected first onset of peaks in the second probe.}
		\label{fig:c4_key2disp_result}
	\end{figure}

\begin{figure}[htbp]
		\centering
		\captionsetup{justification=centering}
			\subfloat[Keypress Behavior - feedback code execution LED(Key2Led), Unity 2019.4.20f]{
				\includegraphics[width=.45\textwidth]{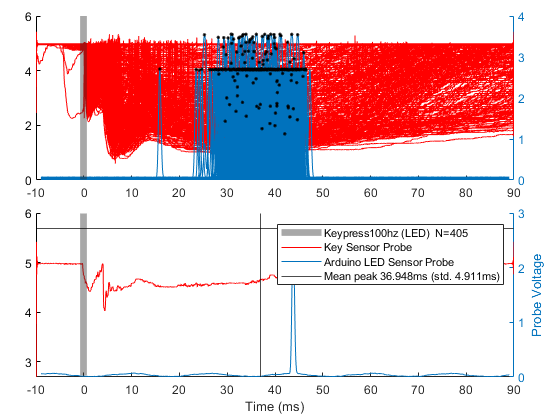}
				\label{subf-c4-key2led100-unity}
			}
			
			\subfloat[Keypress Behavior - feedback code execution LED(Key2Led), Unreal Engine 4.26]{
				\includegraphics[width=.45\textwidth]{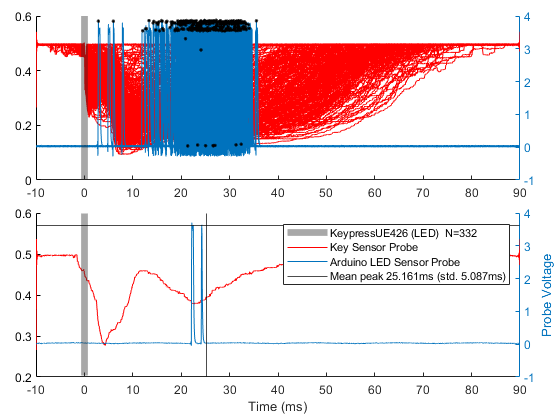}
				\label{subf-c4-key2led-unreal}
			}		
		\caption{Latency measured between keyboard press behavior and marker execution for keyboard event recognition code in 3D engine (Key2Led), for both Unity and Unreal Engine. Higher plot in each set of plots shows all measurements from the two probes superimposed, along with the delayed response peaks determined by Matlab's findpeak() function. Lower plot shows the average waveforms, as well as the average of the automatically detected first onset of peaks in the second probe.}
		\label{fig:c4_key2led_result}
	\end{figure}

From stimulus onset marker code execution to the actual onset of the chess-board grid on the HMD pixels, on Unity Engine there was an average latency of 10.777ms (SD 0.672), while on Unreal Engine an average latency of 21.059ms (SD 0.671) was observed. From behavioral keypress onset detection to chess board grid onset on the HMD, an average latency of 47.026ms (SD 6.156) was observed on Unity while 46.682ms (SD 4.499) was observed on Unreal Engine. In a separate session measuring latency between physical keypress detection and LED onset upon keypress register in the 3D engine, we found an average latency of 36.948ms (SD 4.911) on Unity and 25.161ms (SD 5.087) on Unreal. Table~\ref{tab:c4_vr_measurement_count} shows the summarized results. Figure~\ref{fig:c4_stim2disp_result} (Stim2Disp), ~\ref{fig:c4_key2disp_result} (Key2Disp), and ~\ref{fig:c4_key2led_result} (Key2Led) each shows probe measurements for all individual trials superimposed on the top plot (as well as the detected response peaks as black scatterplots), and the averaged out measurements on the lower plot.
	
	\begin{table*}
		\scriptsize	
		\captionsetup{justification=centering}
		\centering
		\begin{center}
				\begin{tabular}{|l|c|c|}
					\hline	
					Mean latency in ms ($\pm$SD)&\textbf{Unity 2019.4.20f}&\textbf{Unreal Engine 4.26} \\
					\hline
					\textbf{Stim2Disp}&10.777($\pm$0.672)&21.059($\pm$0.671)\\\hline
					\textbf{Key2Disp}&47.026($\pm$6.156)&46.682($\pm$4.499)\\\hline
					\textbf{Key2Led}&36.948($\pm$4.911)&25.161($\pm$5.087)\\\hline
					
					\hline	
					
				\end{tabular}
		\end{center}
		
		\caption{Average and standard deviation values for the measured latency in all conditions, for both 3D engines shown in milliseconds.} 
		\label{tab:c4_vr_measurement_count}	
		
	\end{table*}
	
Two-sample T-tests between the two 3D engines for the Stim2Disp condition showed a significant difference in latency between Unity and Unreal ($t_{df=641}=60.537, p<1e^{-100}, SD=0.665$).
Similarly, a significant difference was observed between Unity and Unreal's latecny in the Key2Led condition ($t_{df´735}=31.900, p<1e^{-100}, SD=4.991$).
No significant difference was found in latency between Unity and Unreal for the Key2Disp condition ($t_{df=713}=0.833, p=0.405, SD=5.484$).

\section{Discussion}
This study aimed to make precise measurements of latencies that may occur during time-critical cognitive behavioral experiments in 3D engine-based virtual reality environments. To achieve this, we implemented a bare-bone 3D environment in both Unity and Unreal Engine, two prominent 3D engines that are used to develop Virtual reality and 3D scenarios for general purposes. We implemented latency measurement in three different scenarios: a scenario in which a marker event for stimuli presentation was sent and rendered on the display, a scenario in which a physical key press by a participant happened and registered in the 3D engine, and a scenario where a key press happened and the resultant feedback occurred on the display. We used oscilloscope probes combined with photodiodes on display events, serial-communication triggered LEDs for software events, and pressure sensors for physical keypress events in order to make precise timing measurements on when each of these events were occurring.

We first discuss the difference in average latency values between Key2Disp and Key2Led conditions: although measurements for the two conditions were made separately (as the Key2Led involved sending a LED trigger to arduino upon 3D engine recognition of the key event), considering our experiments were designed to be as basic as possible in terms of code implementation (as well as the sufficient sample size for each condition), at a glance one would expect the sum of Stim2Disp and Key2Led conditions to match up or be somewhat less than the average Key2Disp values. In Unity, the sum of the average of two conditions exceed the Key2Disp slightly. We believe this is understandable considering the communication time between PC software and the actual Arduino interface itself. In a previous study by Schubert et al.~\cite{schubert2013using}, downstream communication from an experimental computer to an Arduino was measured to be an average of 1.251ms with a low standard deviation. Considering the difference between the sum of mean Key2Led and Stim2Disp, and the Key2Disp condition itself, the difference appears to be enough to explain the somewhat larger mean latency in the combined latency.

\subsection{Registering behavioral response without Unity or Unreal with LSL}
From our current set of results, it appears the largest issue in maintaining a reliable latency for cognitive experiment occurs from registering behavioral responses from the user I/O device. As can be observed from results in the Key2Led condition, in both 3D engines this latency is the largest (and the most variable) in registering the physical response event to the software stack. 

It is possible that the nature of serial port devices contribute a large part in this variance: parallel-port connected devices have been known to be favored over serial port connections for participant I/O in timing critical experimental design ~\cite{beringer1992timing, chambers2003timing, plant2003choice}. However serial port device technology has come a long way, and it has been suggested that the imprecision arising from user input devices may not be as critical as  believed previously~\cite{damian2010does}. In older studies comparing serial and PS/2 devices, serial input devices were reported to have a much higher input latency with high variance ~\cite{chambers2003timing}. In modern devices however, the latency gap between serial port based devices and parallel port devices may have become less considerable: response pad hardware specifically used for cognitive experiments such as Cedrus pads use serial USB connections. Furthermore, the choice of keyboard hardware that was used in our experiment was a mechanical keyboard that uses optic-based switches for faster input recognition on the hardware's part, along with high polling rates over 100Hz.

The software stack also plays as much as a large part in the mean and variation of the latency as the hardware stack does. It has been reported that the experimental software framework as well as the operating system can contribute to differently distributed latency and missed frame counts ~\cite{garaizar2014measuring}. We look into lowering the input variance and latency further by utilizing software stack independent from 3D engines in this section.

\begin{figure}[htbp]
		\centering
		\captionsetup{justification=centering}
			\includegraphics[width=0.45\textwidth]{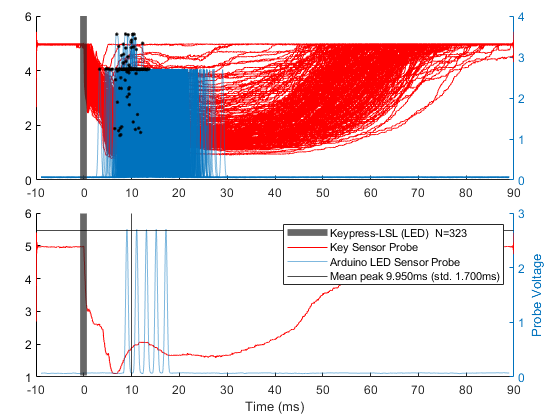}	
		\caption{Keyboard input to C++ LSL code register. Higher plot in each set of plots shows all measurements from the two probes superimposed, along with the delayed response peaks determined by Matlab's findpeak() function. Lower plot shows the average waveforms, as well as the average of the automatically detected first onset of peaks in the second probe.
		}
		\label{fig:c4_key2lsl}
	\end{figure}

In light of these considerations, we performed another set of measurements, but this time using a software outside of 3D engines for key event recognition. Lab Streaming Layer (LSL)~\cite{kothe2014lab} is a C++ based system for synchronizing experimental data from multiple sources through a unified clock (\url{https://github.com/sccn/labstreaminglayer} for more info). LSL has been used for cognitive studies involving physiological signal measurements in which timing was critical~\cite{wang2019assessing, gramann2014imaging, blum2021pocketable, bigdely2013hierarchical, lee2020virtual}. It supports language bindings in multiple programming languages, as well as writing functions for adding custom data source to the data streaming system. We modified a C++ callback code available in the LSL Github Repository to catch certain key events and send Arduino LED events similar to Key2Led condition, but bypassing 3D engines for the key event recognition and getting them directly from the OS level:
	

\begin{figure}
 \begin{lstlisting}[frame=single]
// LSL outlet definition
static lsl::stream_outlet* outlet = nullptr;
// Keyboard hook
static HHOOK kbdHook = nullptr;
static bool isPressed[256] = {0};

//const WCHAR FileFullPath[] = { L"COM4" };
HANDLE hSerial;
char ledcode;
DCB dcbSerialParams = { 0 }; //FILE_ATTRIBUTE_NORMAL
DWORD byteswritten;

LRESULT CALLBACK keyboard_callback
(int code, WPARAM wParam, LPARAM lParam) {
 if (code >= 0 && outlet) {
  unsigned char key = 0;
  switch (wParam) {
   case WM_KEYDOWN:
   case WM_SYSKEYDOWN:
    key = ((KBDLLHOOKSTRUCT *)lParam)->vkCode & 0xFF;
			
    if (!isPressed[key]) {				
     std::string evstr{key_names[key] + " pressed"};
     // push key event to LSL stream
     outlet->push_sample(&evstr);
     std::cout << evstr << std::endl; 
     isPressed[key] = true;
     // send LED trigger to Arduino
     if (!WriteFile(hSerial, &key, 1, &byteswritten, 0))
     {
     	//In case it don't work get comm error and return false
     	std::cout << "Writefile failed" << std::endl;
     }				
   }
   break;
   case WM_KEYUP:
   case WM_SYSKEYUP:
   key = ((KBDLLHOOKSTRUCT *)lParam)->vkCode & 0xFF;
   {
   	std::string evstr{key_names[key] + "released"};
   	// push key release event to LSL
   	outlet->push_sample(&evstr);
   }
   isPressed[key] = false;
   break;
   default:;
   	}
   }
	return CallNextHookEx(kbdHook, code, wParam, lParam);
}

\end{lstlisting}
\caption{Code for sending markers after catching Key events on LSL, and not Unreal or Unity}
\end{figure}	
The results from the set of measurements using LSL and a C++ callback function for key events can be seen in Figure~\ref{fig:c4_key2lsl}. With a mean latency of 9.950ms (SD 1.700) from physical key press event to the Arduino sensor trigger, we are seeing much lower average latency levels that compare to older PS/2 devices, as well as more stable variations in the latency. By logging keypress events or participant behavioral input through separately run programs such as LSL, we believe some of the issues regarding input lag variation in experiments using graphics and compute-heavy 3D engines can be alleviated somewhat.

\subsection{Stimulus presentation code to auditory stimulus onset delay in Unity}
In interactive experiments utilizing VR technology, especially in those that aim to create naturalistic experimental environment with immersion, it is often worth considering a multisensory presentation of stimuli. The addition of auditory components to the visual stimuli presentation would create a much more immersive VR simulation. And like visual stimuli, presentation of auditory stimuli needs to be considerably precise in timing as well for event related design experiments measuring physiological and behavioral response to stimuli.

	\begin{figure}[htbp]
		\centering
		\captionsetup{justification=centering}
			\subfloat[Sound file used for latency measurement]{
				\includegraphics[width=0.45\textwidth]{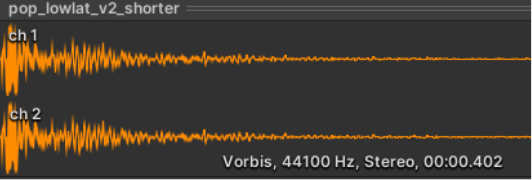}
				\label{subf-c4-soundpic}
			} 
			\\
			\subfloat[Oscilloscope probe setup of sound latency measurement]{
				\includegraphics[width=0.45\textwidth]{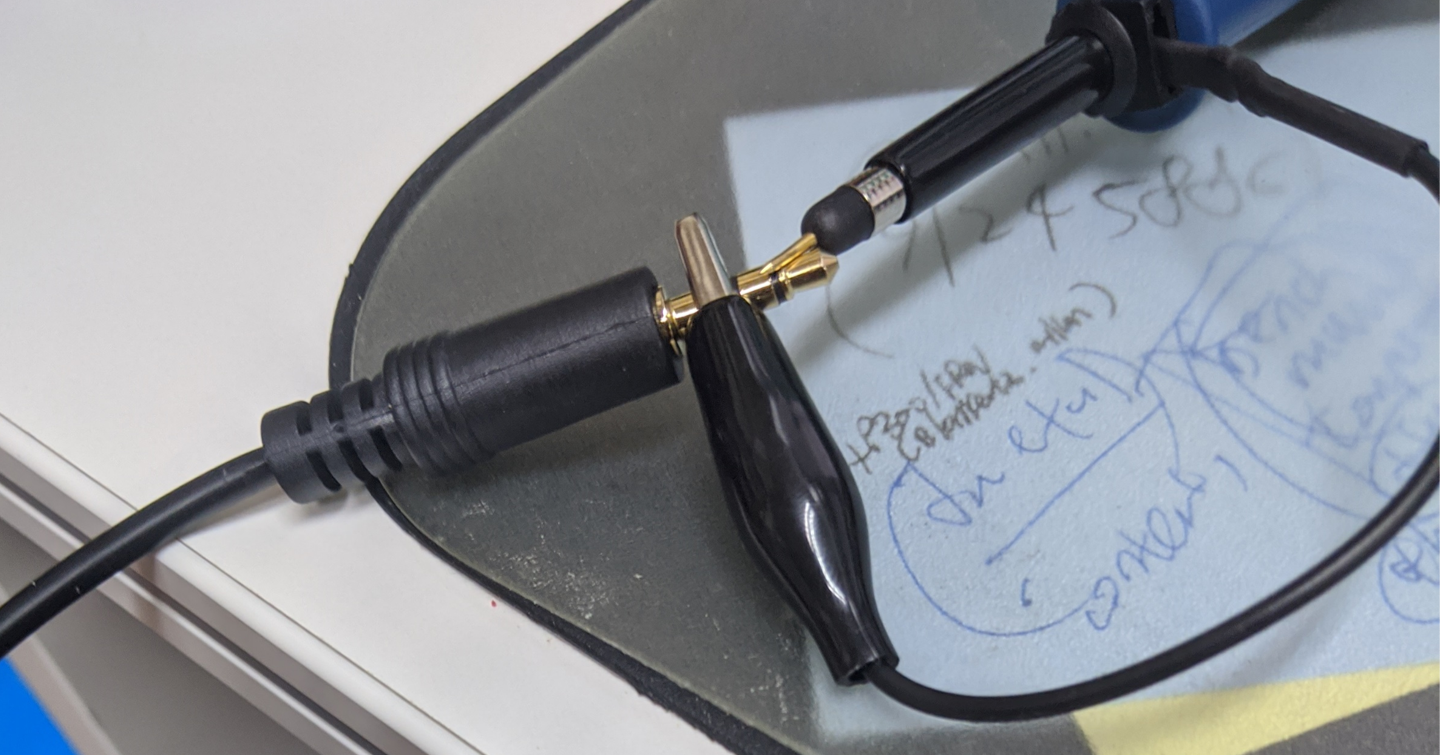}
				\label{subf-c4-soundsetup}
			}		
		\caption{Circuit diagram of the latency measuring apparatus, setup pictures}
		\label{fig:c4_sti2sound_setup}
	\end{figure}
	
\begin{figure}[htbp]
		\centering
		\captionsetup{justification=centering}
			\subfloat[Unity default sound library]{
				\includegraphics[width=0.45\textwidth]{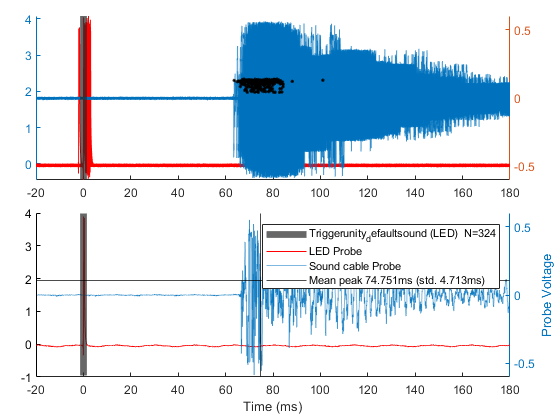}
				\label{subf-c4-stim2sound-b1}
			}
			\\ 
			\subfloat[FMOD Buffer Size 512]{
				\includegraphics[width=0.45\textwidth]{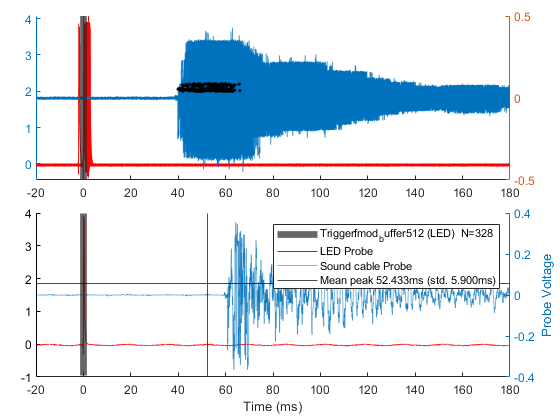}
				\label{subf-c4-stim2sound-b2}
			}		
			\\
			\subfloat[FMOD Buffer Size 1024]{
				\includegraphics[width=0.45\textwidth]{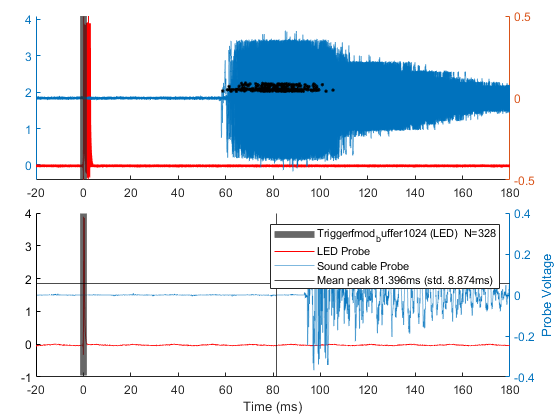}
				\label{subf-c4-stim2sound-b3}
			}	
		
		\caption{Stimuli onset code Led to actual auditory stimuli propagation. Higher plot in each set of plots shows all measurements from the two probes superimposed, along with the delayed response peaks determined by Matlab's findpeak() function. Lower plot shows the average waveforms, as well as the average of the automatically detected first onset of peaks in the second probe.
		}
		\label{fig:c4_stim2sound_result}
	\end{figure}

We deemed it was also worth investigating the latency of auditory stimuli onset code and the physical propagation of the stimuli sound. In the case of Unity, one can utilize the base sound functionality provided by the engine, or, use 3rd-party sound engines that are compatible with the 3D engine such as FMOD (\url{https://www.fmod.com/}). While the default sound library from Unity does not provide a lot of tweaking options to optimize for performance, FMOD allows manually setting sound playback buffer sizes. For this study, we used a soundfile from ~\cite{chen2017pop} as the stimulus to playback, either using Unity's default sound library, or using FMOD with a buffer size of either 512 or 1024. A line-out cable (3.5mm M/M) was plugged into the speaker jack of the experimental computer, with the other end being connected to the oscilloscope probe as can be seen in Figure~\ref{fig:c4_sti2sound_setup}. We did similar measurements like in Stim2Disp conditon, measuring the latency between stimuli onset code and the actual propagation of the sound in the sound cable. We report the results in Figure~\ref{fig:c4_stim2sound_result}. While using FMOD with a buffer size of 512 yielded the best results, we observe that latency for auditory stimuli presentation is much worse compared to visual stimuli both in mean accuracy and in variation. Based on this observation, we believe caution is warranted when using auditory stimuli, especially when in the absence of a concurrent visual stimuli.

	
	
	
\subsection{Future works}
As Wimmer et al.~\cite{wimmer2019latency} reported, specific latency measures for experimental setup are strongly dependent on the specifics of the hardware and software one acquires for the experiment. Considering the continuously developing landscape in VR and its related hardware/software stack, simply measuring the latency of each setup is not only insufficient, but it is a fruitless endeavor long-term. Instead, it would be more prudent to develop a framework capable of measuring delays for configurable setup on the go: this is our most immediate next step. Furthermore, the purpose of establishing latency value distributions are to ultimately utilize them in developments of VR-based behavioral experiments in event-related designs to collect synchronized time-dependent behavioral and physiological data; for our purposes of investigating underlying brain processes, we are especially interested in utilizing these findings to create latency-optimized VR EEG experiments in immersive, naturalistic 3D.


\section*{Acknowledgments}
This study was supported by the National Research Foundation of Korea under project BK21 FOUR and grants NRF-2022R1A2C2092118, NRF-2022R1H1A2092007, NRF-2019R1A2C2007612, as well as by Institute of Information \& Communications Technology Planning \& Evaluation (IITP) grants funded by the Korea government (No. 2017-0-00451, Development of BCI based Brain and Cognitive Computing Technology for Recognizing User’s Intentions using Deep Learning; No. 2019-0-00079, Department of Artificial Intelligence, Korea University; No. 2021-0-02068, Artificial Intelligence Innovation Hub).

\newpage
\bibliographystyle{IEEEtran}
\bibliography{ownpubs}

\vfill

\end{document}